\documentclass[sigconf]{acmart}
\usepackage{graphicx}
\usepackage{amsmath}
\usepackage{multirow}
\usepackage{svg}

\usepackage{amssymb}
\usepackage{xcolor}
\usepackage[T1]{fontenc}
\usepackage{hyperref}
\usepackage{amsfonts}
\usepackage{tabularx}
\usepackage{microtype}
\hypersetup{
  colorlinks   = true, 
  urlcolor     = blue, 
  linkcolor    = blue, 
  citecolor   = blue 
}
\usepackage{cleveref}
\AtBeginDocument{%
  }

\setcopyright{acmlicensed}
\copyrightyear{2025}
\acmYear{2025}
\setcopyright{cc}
\setcctype{by}
\acmConference[SIGIR '25]{Proceedings of the 48th International ACM SIGIR Conference on Research and Development in Information Retrieval}{July 13--18, 2025}{Padua, Italy}
\acmBooktitle{Proceedings of the 48th International ACM SIGIR Conference on Research and Development in Information Retrieval (SIGIR '25), July 13--18, 2025, Padua, Italy}\acmDOI{10.1145/3726302.3730239}
\acmISBN{979-8-4007-1592-1/2025/07}

\begin{document}
\pagestyle{fancy}

\title{Rational Retrieval Acts: Leveraging Pragmatic Reasoning to Improve Sparse Retrieval}

\author{Arthur Satouf}
\email{arthur.satouf(at)gmail.com}
\affiliation{%
  \institution{Université Paris-Saclay, ILLS, ISIR, \\ Air Liquide }
  \country{France, Canada}
}

\author{Gabriel Ben-Zenou}\authornote{The author conducted this work as an intern at ILLS.}
\email{gabriel.ben-zenou(at)polytechnique.edu}
\affiliation{%
  \institution{CentraleSupélec, ILLS, AMIAD}
  \country{France, Canada}
}

\author{Benjamin Piwowarski}
\email{benjamin.piwowarski(at)cnrs.fr}
\affiliation{%
  \institution{Sorbonne Université, ISIR, CNRS}
  \country{France}
}

\author{Habiboulaye Amadou-Boubacar}
\email{habiboulaye.amadou-boubacar(at)airliquide.com}
\affiliation{%
  \institution{Air Liquide}
  \country{France}
}

\author{Pablo Piantanida}
\email{pablo.piantanida(at)cnrs.fr}
\affiliation{%
  \institution{MILA - Quebec AI Institute, Université Paris-Saclay, CentraleSupélec, ILLS, CNRS}
  \country{France, Canada}
}

\renewcommand{\shortauthors}{Arthur Satouf, Gabriel Ben-Zenou,Benjamin Piwowarski, Pablo Piantanida, Habiboulaye Amadou-Boubacar}

\begin{abstract}
 	
Current sparse neural information retrieval (IR) methods, and to a lesser extent more traditional models such as BM25, do not take into account the document collection and the complex interplay between different term weights when representing a single document. In this paper, we show how the Rational Speech Acts (RSA), a linguistics framework used to minimize the number of features to be communicated when identifying an object in a set, can be adapted to the IR case -- and in particular to the high number of potential features (here, tokens). RSA dynamically modulates token-document interactions by considering the influence of other documents in the dataset,  better contrasting document representations. Experiments show that incorporating RSA consistently improves multiple sparse retrieval models and achieves state-of-the-art performance on out-of-domain datasets from the BEIR benchmark. https://github.com/arthur-75/Rational-Retrieval-Acts

\end{abstract}



\received{23 January 2025}
\received[revised]{23 January 2025}
\received[accepted]{4 April 2025}

\maketitle
\section{Introduction}
The IR field evolved rapidly since the transformer architecture has been adopted to design various types of models, ranging from effective but less efficient cross-encoders~\cite{nogueira2019multistagedocumentrankingbert} to more efficient models based on dual architectures \cite{hofstätter2021efficientlyteachingeffectivedense,ref_spalde}. Among the latter, sparse neural IR models like e.g. SPLADE~\cite{ref_spalde} are state-of-the-art with a performance well above that of standard IR models like BM25~\cite{bm25} on most datasets. However, on the document representation side, \cite{MackenzieWackyWeightsLearned2021} showed that transformer-based sparse IR model can produce  “wacky weights” -- i.e., term weights with limited discriminative power. 
This issue is compounded by the typically small number of query keywords, as well as the relatively low contrastive ability of existing models. Even traditional models such as BM25 do not fully capture the interplay among terms within documents.

Another issue with sparse representations is on the query (user) side. Azzopardi and Zuccon~\cite{AzzopardiBuildingEconomicModels2019} describe users as rational agents when they interact with a search engine: they seek to reduce the number of keywords they type. 
Such economy-based theories in IR echo the one of Grice in linguistics~\cite{Grice}, that focuses on how people convey meaning beyond what is explicitly stated. These ideas have been mathematically formalized in the Rational Speech Acts (RSA) framework, proposed by Frank and Goodman \cite{RSA_original_paper}. RSA models human-like communication between a speaker and a listener by assuming a shared common knowledge among pragmatic agents that helps lower the communication cost (number of utterances) while keeping meaning unambiguous.

While users behave as pragmatic speakers~\cite{AzzopardiBuildingEconomicModels2019}, existing IR models are not pragmatic listeners~\cite{MackenzieWackyWeightsLearned2021} -- and we argue here that this discrepancy can be reduced using the RSA framework, especially for out-of-domain adaptation. More precisely, we show that applying RSA to IR, to transform literal document representations to pragmatic ones, improves the performance of  IR sparse models. 
Neural networks and modern IR models already capture meaning beyond individual token semantics, for example through attention mechanisms, but, in addition, we distinguish here between \emph{literal} representations (before RSA is applied) and \emph{pragmatic} representations (after RSA refines them).
The word \emph{pragmatic} here is to be understood in the sense of Bunt's concept of pragmatic behavior \cite{Bunt}, that is acting in consideration of the context. RSA incorporates such context with the consideration of external information (in our case, all possibly queried documents and all tokens of the vocabulary). 

Cohn-Gordon et al. \cite{CohnGordon2018, CohnGordon2019} 
first applied RSA to image captioning and machine translation to reduce ambiguity.
In image captioning, RSA refines the output of a language generation model, ensuring the generated caption distinguishes the target image from others in the dataset. 
In translation, RSA modifies the English translation of a German sentence to ensure unambiguous meaning relative to other sentences in the set of sentences to translate. In both cases, the goal is produce the most suitable utterance possible regarding the context -- which is the set of all possible meanings to be conveyed, i.e., still in Bunt's sense \cite{Bunt}, the most pragmatic utterances. In other terms, RSA brings models up to humans pragmatic level.
Just as in language tasks, we show in this paper that RSA enhances IR by making document representations more pragmatic.

RSA has also been employed to model various linguistic phenomena \cite{Degen2023,GoodmanFrank2016,Degen2015,frank2016rational}, and its simplicity and explainability make it a powerful tool for modeling human communicative behavior. However, this simplicity also poses challenges when scaling the RSA  to real-world scenarios with significantly larger datasets. To date, only a few works have successfully integrated RSA with AI-driven language models, often requiring adaptations to the RSA mechanism itself~ \cite{CohnGordon2018,CohnGordon2019,shenPragmaticallyInformativeText2019,kimWillSoundMe2020,kimPerspectivetakingPragmaticsGenerating2021}. Similarly, applying RSA to IR requires adjustments to handle large lexicons, leveraging sparse IR model outputs to our advantage.

In this paper, we make the following contributions:
\begin{itemize}
    \item We introduce Rational Retrieval Acts (RRA), an adaptation of the RSA ~\cite{RSA_original_paper} for sparse IR models, carefully extending it to handle a large number of documents (meanings) and tokens (utterances).
    \item We achieve significant performance improvements in out-of-domain sparse retrieval without increasing inference costs.
\end{itemize}

\section{The Rational Retrieval Acts (RRA)}
To apply the RSA framework to IR, we map tokens from the vocabulary 
\(\mathcal{T}\) to RSA's \emph{utterances} and documents from the set \(\mathcal{D}\) to RSA's \emph{meanings}. For clarity, we use the IR terminology (tokens, documents) instead of the RSA's (utterances, meaning).

\subsection{Integrating the RSA}\label{subsec:Integrating the RSA}
RSA initially assumes that we can define a positive weight \(\mathcal{L}(t,d) \in \mathbb{R}^+\) for each token-document pair $(t,d)\in \mathcal T \times \mathcal D$. The function $\mathcal{L}$ is called the Initial Lexicon. In this paper, we propose to derive \(\mathcal{L}(t,d)\) directly from the weights \(w_{t,d}\) predicted by a sparse IR model using the following equation:
\begin{equation}
\label{eq:w2L} 
\mathcal{L}(t,d) = f(w_{t,d})\ ,
\end{equation}
where \(f\) represents a well chosen initial transformation function.

RSA then defines the literal listener distribution over the documents conditioned on each token, $L_0(d|t)$. 
It represents how the IR model interprets $t$, indicating how likely each document $d$ is to be relevant if $t$ appears in the query.
 The formula defining $L_0$ also includes a prior term $\mathbf{P}(d)$, which can serve the same role as in probabilistic approaches in information retrieval, e.g.\cite{Pontelanguagemodelingapproach1998}. The exact formulation given by RSA is: 
\begin{equation}
    L_0(d|t) = \frac{\mathbf{P}(d) \cdot \mathcal{L}(t,d)}{\sum_{d'\in \mathcal{D}}\mathbf{P}(d') \cdot \mathcal{L}(t,d')}\ , \label{eq:L0}
\end{equation}
where we can write the normalizing factor dependent on $t$ as $Z_t^{(0)}=\sum_{d'\in \mathcal{D}}\mathbf{P}(d') \cdot \mathcal{L}(t,d')$.

When no prior knowledge about the documents is available, P(d) is constant across all documents, and the literal listener simplifies to the term-normalized weights  $w_{t,d}$ from the retrieval model.

Pragmatism is then introduced through the pragmatic speaker $S_1(t|d)$, a distribution over tokens conditioned on documents. 
This distribution models the likelihood of a pragmatic user selecting a token to describe a given document, assuming they interpret token-document associations similarly to the retrieval model. In other words, users align with the model's learned representation of term-document relationships but refine their choice of tokens based on the whole sets of documents and tokens.
RSA introduces a hyper-parameter $\alpha$ to control how pragmatic the user is.
We discuss our choice of $\alpha$ in section \ref{subsec:setup}. The RSA defines $S_1$ as:
\begin{equation}\label{eq:speaker} 
    S_1(t|d)  =  \frac{\exp (\alpha \cdot \log(L_{0}(d|t)))}{Z_d^{(1)}} \ , 
\end{equation} 
where $Z_d^{(1)}$ is the normalizing factor for a given $d$.

If the pragmatic speaker is a model of the user's behavior when formulating a query, what interests us is how the retrieval information system should react to this human pragmatic behavior. In the RSA, such reaction is modeled with the pragmatic listener $L_1(d|t)$. Its definition in the RSA is the following:
\begin{align}\label{eq:listener}
    L_1(d|t) =   \frac{\mathbf{P}(d) \cdot S_1(t|d)}{Z_t^{(1)}}\ ,   
\end{align}
where $Z_t^{(1)}$ is the normalizing factor for a given $t$.

One could apply RSA in an iterative way, computing a new pragmatic speaker-listener pair on top of the last pragmatic listener. However, previous works on RSA choose not to \cite{CohnGordon2018,CohnGordon2019}, especially as the $\alpha$ parameter allows already to balance the level of pragmatic depth of RSA.   




\begin{figure}
  \centering
  \includegraphics[width=\columnwidth]{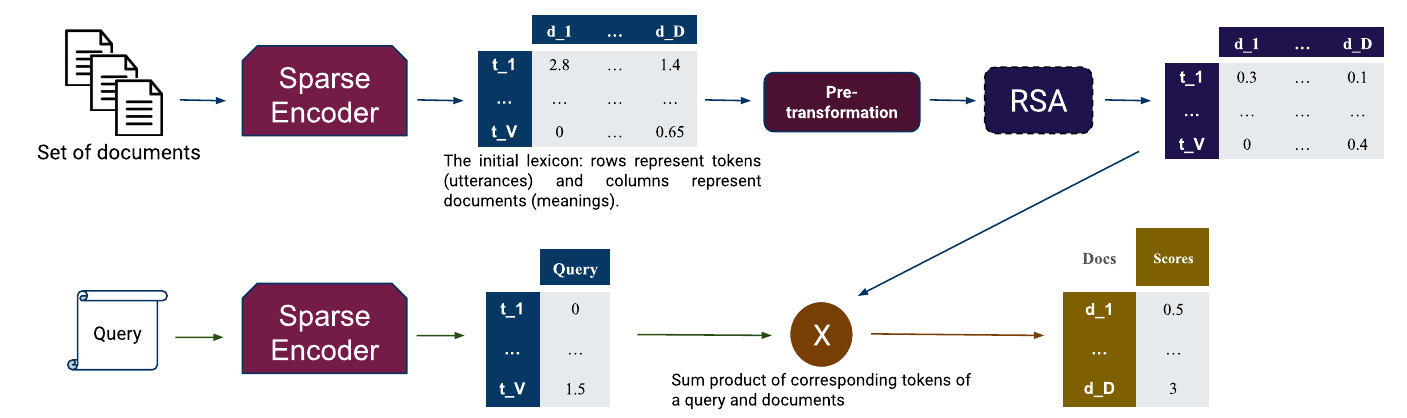}
  \caption{Rational Retrieval Acts (RRA) architecture. Document relevance $L_1(d|t)$ is computed via RSA, assuming a pragmatic user (\emph{speaker}) query. Final ranking uses the dot product between query and document representations (Eq.\eqref{eq:rsv})}
  \label{fig:RRA_diagram}
\end{figure}
  
\subsection{Leveraging the sparsity in IR}\label{subsec:Leveraging the sparsity in IR}
Naively computing speakers' and listeners' distributions with equations \eqref{eq:speaker} and \eqref{eq:listener} is not possible in IR regarding memory resources given the large number of documents and terms (both $S_1$ and $L_1$ matrices are of size $|\mathcal{T}|\times|\mathcal{D}|$). To overcome this dimensionality issue, one way is to reduce the size of the possible meanings by excluding documents that are graded with a low similarity score with the query \cite{andreasReasoningPragmaticsNeural2016,monroeColorsContextPragmatic2017}. However, such shrinking method collides with our need to run a fixed RSA for the whole database at once. Cohn Gordon et al. \cite{CohnGordon2018,CohnGordon2019} proceed by leveraging the sequential aspect of generated text to make the space of utterances smaller. 

Similarly, we propose a technique that relies on the \emph{sparsity} of the sparse IR models we apply RSA to. Such sparsity means that most of the tokens of the vocabulary do not occur in each document i.e. for most $(t,d)\in\mathcal{T}\times\mathcal{D}$, $w_{t,d}=0$. The application of the non linear transformation function $f$ posterior to the sparse models thus lead most entries of the RSA $\mathcal{L}(t,d)$ to be equal to $f(0)$. We show in the following that decomposing $\mathcal{L}(t,d)$ into two factors $l_t^{(0)}$ and $l_d^{(0)}$ corresponding respectively to the global importance of a term and a document for all $(t,d)$ pair such that $w_{t,d}=0$ allows to drastically reduce the required memory storage from $|\mathcal{T}|\times |\mathcal{D}|$ to $|\mathcal{T}| + |\mathcal{D}|$ to which the varying (small) size of the sparse output weight tensor of the IR model needs to be added. 

Formally, we first denote $\mathcal{D}_t$ the set of documents where the token $t$ has a non null weight ($w_{t,d} \not=0$), and similarly, $\mathcal{T}_d$ the set of tokens which have a non-null weight for document $d$.
Then, for any token $t$ and document $d$ such as $d \not\in \mathcal D_t$
\begin{equation}
    L_0(d\mid t) = \underbrace{\frac{\mathbf{P}(d)\cdot f(0)}{\sum_{d' \in \mathcal{D}_t} \mathbf{P}(d')\cdot  f(w_{t,d'}) + \sum_{d' \notin \mathcal{D}_t} \mathbf{P}(d')\cdot f(0)}}_{l_t^{(0)}} \times \underbrace{1\vphantom{\frac {z}{\sum_{d'\in \mathcal D_t}}}}_{l_d^{(0)}}\ . 
    \label{eq:L0:sparse}
\end{equation}


Leveraging this initial decomposition, we can then rewrite the equations \eqref{eq:speaker}  and \eqref{eq:listener} for any token $t$ and document $d$ such that $d \not\in \mathcal D_t$.
First, for the pragmatic speaker, we can rewrite equation \eqref{eq:speaker} $\forall  t \not\in \mathcal T_d$:
\begin{equation}
    S_1(t\mid d)  = 
    \underbrace{
    {\exp\left(\alpha \log l_t^{(0)} \right)}}_{s_t^{(1)}}
    \times 
    \underbrace{ 
    \exp\left(\alpha \log l_d^{(0)} \right)
    / {Z_d^{(1)}}}_{s_d^{(1)}}, \label{eq:speaker:sparse}
\end{equation}
where the normalizing factor $Z^{(1)}_d$ is computed as
\begin{equation}
    Z_d^{(1)} = \sum_{t \in \mathcal{T}_d} \exp \left( \alpha L_0(d|t) \right) + \sum_{t \not\in \mathcal{T}_d} \exp\left(\alpha \log (l_{t}^{(0)} l_d^{(0)}) \right). \label{eq:speaker:normalization}
\end{equation}

Based on this reformulation of the pragmatic speaker, we can rewrite the pragmatic listener \eqref{eq:listener} for all  $d \not\in \mathcal D_t$:
\begin{eqnarray}
    L_1(d\mid t) = 
    \underbrace{{s_t^{(1)}} / {Z_t^{(1)}}}_{l_t^{(1)}} 
    \times 
    \underbrace{s_d^{(1)}}_{l_d^{(1)}}, \label{eq:listener:sparse}
\end{eqnarray}
with the normalizing factor $Z_t^{(1)}$ computed as
\begin{equation}
    Z_t^{(1)} = \sum_{d' \in \mathcal{D}_t} S_1(d|t) + \sum_{d' \notin \mathcal{D}_t} s_t^{(1)}s_d^{(1)}. 
    \label{eq:listener:normalization}
\end{equation}

Let us insist on the fact that this does not change at all the RSA process -- this reformulation work is made for computing complexity reasons, and is necessary when applying RRA.
\subsection{Scoring documents}\label{subsec:scoring documents}
The final representation of a document after applying the RSA is $(L_1(d|t))_{t\in\mathcal{T}}$. While RSA alternation between a speaker and a listener can theoretically be run for several iterations, practical applications of Rational Speech Acts (RSA) modeling to human linguistic behavior typically restrict it to a single iteration. \citet{frank2016rational} shows that $\alpha$ and the number of iterations balance each other out. We thus fix the number of iterations to 1 and detail our choice of $\alpha$ in section \ref{subsec:setup}.

Finally, as depicted in Figure \ref{fig:RRA_diagram}, following RSA, the score of a given document is given by  
\begin{equation}
\mathrm{score}(q,d) = \sum_{t \in \mathcal T_d} w_{t,q}\times L_1(d|t) + l_d^{(1)} \times \sum_{t \notin \mathcal T_d} w_{t,q}\times l_t^{(1)} , \label{eq:rsv}
\end{equation}
where $w_{t,q}$ is the value for token $t$ of the representation of the query on the IR model vocabulary. 

All together, equations \eqref{eq:L0:sparse} to \eqref{eq:listener:normalization} define the computation of the document representations in our Rational Retrieval Acts (RRA) framework. Compared to the original RSA equations (\eqref{eq:L0} to \eqref{eq:listener}), this allows for an memory efficient computation of the pragmatic listeners and speakers. At retrieval time, equation \eqref{eq:rsv} allows to leverage inverted indices for fast retrieval, with minor adaptations. RRA can thus be applied to large-scale document collections. 

\section{Experiments}
\subsection{Setup}\label{subsec:setup} 
For our experiments, we implement the RRA pipeline illustrated by Figure \ref{fig:RRA_diagram}. The chosen sparse encoders and datasets of documents and queries are detailed in subsection \ref{subsec:Baselines and datasets}. Our choice of pre-transformation function is explained in subsection \ref{subsec:Pre-transformation Function}. RSA requires both a prior and a $\alpha$ parameter to be defined. For the prior, we assume it is uniform over the documents as no specific prior knowledge is given. It translates into $\mathbf{P}(d)=\frac{1}{|\mathcal{D}|}$ for all $d\in\mathcal{D}$. 

\begin{figure}
  \centering
  \includegraphics[width=\columnwidth]{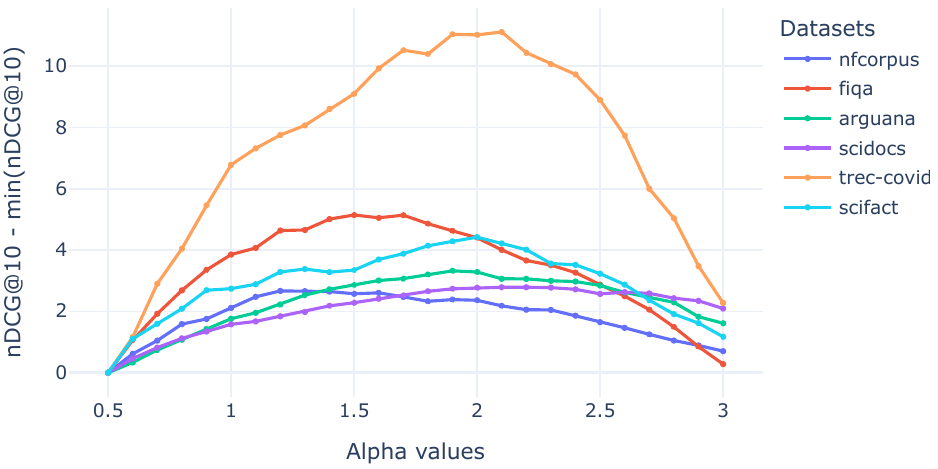}
  \caption{Impact of $\alpha$ on S-RRA performance (nDCG@10). Non-optimal $\alpha$ values can significantly degrade results.}
  \label{fig:alphas_evolution}
\end{figure}

Regarding $\alpha$, as its value tends to infinity, equation \ref{eq:speaker} makes it clear that the probability density function given by the pragmatic speaker tends to a Dirac. Conversely, as alpha goes to 0, it tends towards a uniform density. In the first case, only one document is given a non-zero probability (equal to 1), making it impossible to distinguish between subsequent documents; in the second case, documents all have the same importance, and thus cannot be distinguished. Prior work in RSA suggests that $\alpha = 1$ gives the best fit between simulated and human behavior. However, in practice, the choice of $\alpha$ may depend on dataset characteristics and retrieval model parameters. Figure \ref{fig:alphas_evolution} shows the behavior of the S-RRA (Splade+RSA) model for different values of $\alpha$: we notice a difference of performance reaching up to 10 points of nDCG@10 on the TREC-COVID dataset depending on the choice of the $\alpha$ parameter. Therefore, it is expected that performance is bounded between these two extremes and that an optimal alpha exists. The performance is however quite stable when varying $\alpha$, and we can predict a close-to-optimal value of $\alpha$ by validating $\alpha$ on few synthetic queries for each dataset. We follow the methodology in \cite{BonifacioInParsUnsupervisedDataset2022},
and sample  500 documents from a given dataset. For each document, we use LLAMA3-8B \cite{dubey2024llama3herdmodels} to generate a single, relevant query for each document. Specifically, we prompt the model to create a diverse set of queries. 
This process produces a synthetic information retrieval (IR) dataset consisting of query-document pairs that are considered relevant. This in turns allows us to select the $\alpha$ that maximizes nDCG@10 on the synthetic dataset. Note that this dataset is relatively small, and experiments using the same queries to fine-tune a neural IR sparse model did not give any positive results.
\paragraph{Pre-transformation function}\label{subsec:Pre-transformation Function}
Finally, preliminary experiments on MS-Marco have shown that the pre-transformation function $f:x\rightarrow 1+x$ was performing the best among the following mappings: $f:x\rightarrow x$, $\log(1+x)$, $\exp{(x)}$, $1+x$, $\lambda \cdot x$, and $\tanh{(x)}$.
We also observed that functions verifying $f(0) \neq 0$ give better results than the rest, likely because this avoids contrasting too much the document weights $L_1(t|d)$.

\subsection{Baselines and datasets}\label{subsec:Baselines and datasets} 

Given its strong performance, we adopt SPLADE~\cite{ref_spalde} (specifically \texttt{splade-v3}\footnote{\url{https://huggingface.co/naver/splade-v3}}) as our main IR model. We evaluate on the BEIR benchmark~\cite{thakur2021beirheterogenousbenchmarkzeroshot} in a zero-shot setting and, after verifying SPLADE’s effectiveness, 
apply RSA in the same manner to other sparse models (SPARTA~\cite{zhao2020spartaefficientopendomainquestion}, BM25, Unicoil~\cite{lin2021briefnotesdeepimpactcoil}, DeepImpact~\cite{mallia2021learningpassageimpactsinverted}). 
Each model is compared with its RSA-enhanced version (e.g., SPLADE vs. SPLADE+RSA (S-RRA) and SPARTA vs. SPARTA+RSA (SP-RRA)), as shown in Table \ref{tab:rsa_ndcg_results}. To implement these methods, we rely on open-source tools such as Anserini, BEIR, and the models’ original codebases.
We also report the results in Table \ref{tab:rsa_ndcg_results9} of SparseEmbed~\cite{SparseEmbed}, ColBERT-v2 \cite{santhanam2022colbertv2effectiveefficientretrieval}, as well as the performance of BM25 -- alone, with a cross-encoder or used with document expansion from docT5query(D2Q)~\cite{Cheriton2019FromDT} that we copy from~\cite{wang2020minilmdeepselfattentiondistillation}.

\subsection{Results}\label{subsubsec:Results}


\begin{table}[h]
    \centering
    \renewcommand{\arraystretch}{1}
    \footnotesize
    \captionsetup{font=small}
    \caption{Table 1: nDCG@10 across datasets for sparse models and their RSA-enhanced variants (-RRA). Bold: best. Underlined: significant (p<0.05, paired t-test).}
    \label{tab:rsa_ndcg_results}
    \begin{tabular}{p{1.3cm}|p{.4cm}p{.3cm}|p{.4cm}p{.3cm}|p{.3cm}p{.3cm}|p{.5cm}p{.3cm}|p{.4cm}p{.3cm}}
        \toprule
        \textbf{nDCG@10} & Splade & S-RRA & Sparta & SP-RRA & bm25 & bm-RRA & Unicoil & U-RRA & Deep-Impact & D-RRA \\
        \midrule
        ArguAna & 48.9 & \underline{\textbf{50.4}} & 37.1 & \underline{\textbf{44.1}} & 34.8 & \underline{\textbf{38.4}} & 37.4 & \underline{\textbf{41.1}} & {31.8} & \underline{\textbf{41.4}} \\ 
        FiQA-2018 & 37.8 & \textbf{38.3} & 19.8 & \underline{\textbf{24.1}} & 23.8 & \textbf{23.9} & 28.9 & \textbf{29.5} & 27.8 & \underline{\textbf{29.2}} \\
        NFCorpus & 36.4 & \textbf{36.6} & 30.1 & \textbf{30.7} & 32.5 & \textbf{32.7} & 33.3 & \textbf{33.6} & 27.1 & \underline{\textbf{27.8}} \\ 
        Quora & 81.4 & \underline{\textbf{84.0}} & 62.5 & \underline{\textbf{71.5}} & 78.9 & \textbf{79.1} & 66.5 & \underline{\textbf{77.2}} & 66.0 & \underline{\textbf{68.0}} \\ 
        SCIDOCS & 15.5 & \underline{\textbf{16.6}} & 12.6 &\underline{\textbf{14.4}} & 15.8 & {\textbf{15.9}} & 14.4 & \underline{\textbf{15.9}} & 13.4 & \underline{\textbf{14.6}} \\ 
        SciFact & 71.6 & \underline{\textbf{73.2}} & 59.8 & \underline{\textbf{62.4}} & 66.2 & {\textbf{67.2}} & 68.6 & \textbf{69.1} & 59.3 & \underline{\textbf{61.9}} \\ 
        TREC-Covid & 74.2 & \textbf{74.7} & 53.8 & \underline{\textbf{59.4}} & 65.3 & {\textbf{66.8}} & 64.4 & \textbf{65.7} & 56.2 & \underline{\textbf{62.2}} \\
        Touch2020 & 32.2 & \textbf{32.3} & 17.7 & \underline{\textbf{21.6}} & 36.7 & \textbf{37.2} & 29.1 & \textbf{30.0} & 26.0 & \textbf{26.8} \\ 
        CQADupStack & 34.5 & \underline{\textbf{35.7}} & 27.5 & \underline{\textbf{30.4}} & 30.0 & \underline{\textbf{31.3}} & 30.2 & \underline{\textbf{31.2}} & 28.5 & \underline{\textbf{30.4}} \\
        \midrule
        \textbf{Mean} & 48.1 & \textbf{49.1} & 35.7 & \textbf{39.8} & 42.7 & \textbf{43.6} & 41.4 & \textbf{43.7} & 37.3 & \textbf{40.3} \\
        \textbf{Point Gain} & & \textbf{+1} & & \textbf{+4.2} & & \textbf{+0.9} & & \textbf{+2.3} & & \textbf{+2.9} \\
        \bottomrule
    \end{tabular}
\end{table}
Our experimental results~(Table~\ref{tab:rsa_ndcg_results}) highlight RSA’s effectiveness (“*-RRA” columns) in enhancing sparse retrieval models across BEIR datasets. The impact is particularly notable in datasets with longer queries relative to document lengths, such as ArguAna (192.98 query words vs. 166.80 document words, ratio = 1.2) and Quora (ratio = 0.83), where RSA significantly improves performance. Conversely, for datasets like NFCorpus (ratio = 0.01) and Touché-2020 (ratio = 0.02), where queries are much shorter, the benefits of RSA are less pronounced.
BM25, which already captures global term interplay through IDF, benefits less from RSA since its term-weighting mechanism is inherently corpus-aware. Unlike Splade and Unicoil, Sparta and DeepImpact do use a uniform query token weight and see substantial gains with RSA. By dynamically adjusting token importance, RSA compensates for their lack of contextual query term weighting, enhancing their ability to differentiate between documents effectively. As shown in Table~\ref{tab:rsa_ndcg_results9}, integrating RSA into SPLADE (S-RRA) improves performance on most datasets, with an average nDCG@10 increase of 1 point. We hypothesize that RSA refines token-document interactions, leading to better document contrast. However, its impact is less pronounced on datasets such as NFCorpus, DBPedia, and Touché-2020, where modulating token-document interactions plays a smaller role in retrieval effectiveness.

\begin{table}[h]
    \centering
    \renewcommand{\arraystretch}{1}
    \footnotesize
    \captionsetup{font=small}
    \caption{nDCG@10 performance comparison. Our model S-RRA vs state-of-the-art Bold (highest score).}
    \label{tab:rsa_ndcg_results9}
    \begin{tabular}{p{1.2cm}p{.4cm}p{.4cm}p{.4cm}p{.4cm}p{.4cm}p{.4cm}p{.4cm}p{.4cm}}
        \toprule
        \textbf{nDCG@10} & BM25 & Col\allowbreak{}BERT\allowbreak{}v2 & BM25\allowbreak{}+CE & TAS-B & D2Q & Sparse\allowbreak{}Embed & Splade & s-RRA \\
        \midrule
        ArguAna & 31.5 & 46.3 & 31.1 & 42.9 & 34.9 & \textbf{51.2} & 48.9 & 50.4 \\
        CimateFever & 21.3 & 17.6 & 25.3 & 22.8 & 20.1 & 21.8 & 25.0 & \textbf{27.5} \\
        DBPedia & 31.3 & 44.6 & 40.9 & 38.4 & 33.1 & \textbf{45.7} & 44.3 & 44.4 \\
        Fever & 75.3 & 78.5 & \textbf{81.9} & 70.0 & 71.4 & 79.6 & 80.5 & 81.4 \\
        FiQA-2018 & 23.6 & 35.6 & 34.7 & 30.0 & 29.1 & 33.5 & 37.8 & \textbf{38.3} \\
        HotpotQA & 60.3 & 66.7 & \textbf{70.7} & 58.4 & 58.0 & 69.7 & 69.5 & 70.3\\
        NFCorpus & 32.5 & 33.8 & 35.0 & 31.9 & 32.8 & 34.1 & 36.4 & \textbf{36.6} \\
        NQ & 32.9 & 56.2 & 53.3 & 46.3 & 39.9 & 54.4 & 58.3 & \textbf{58.9} \\
        Quora & 78.9 & 85.2 & 82.5 & \textbf{83.5} & 80.2 & 84.9 & 81.4 & 84.0 \\
        SCIDOCS & 15.8 & 15.4 & 16.6 & 14.9 & 16.2 & 16.0 & 15.5 & \textbf{16.6} \\
        SciFact & 66.5 & 69.3 & 68.8 & 64.3 & 67.5 & 70.6 & 71.6 & \textbf{73.2} \\
        TREC-Covid & 65.6 & 73.8 & 75.7 & 48.1 & 71.3 & 72.4 & 74.2 & \textbf{74.7} \\
        Touch2020 & 36.7 & 26.3 & 27.1 & 16.2 & 34.7 & 27.3 & 32.2 & \textbf{32.3} \\
        \midrule
        \textbf{Mean} & 44.0 & 49.9 & 49.5 & 43.7 & 45.3 & 50.9 & 52.0 & \textbf{53.0} \\
        \bottomrule
    \end{tabular}
\end{table}

\paragraph{Case Study: SPARTA on SciFact.}

In the SciFact dataset using the SPARTA model, we observe that without RSA, the token \textbf{“and”}—and other common stopwords—can receive disproportionately high weights. For example, in the first document, \textbf{“and”} is assigned a weight of 1.2515, which is higher than more informative tokens such as \textbf{“statistics”} 1.1672 and \textbf{“evolution”} 1.0842. After applying RSA, this pattern is reversed: \textbf{“and”} is significantly down-weighted to 0.00196, while \textbf{“statistics”} and \textbf{“evolution”} are promoted to 0.00633 and 0.00467 respectively—making \textbf{“and”} roughly three times less prominent. This example highlights how RSA suppresses frequently occurring, less informative tokens across the collection, while amplifying the weights of more distinctive, content-rich tokens. This dynamic adjustment improves the contrastiveness and discriminative power of sparse representations.

\section{Discussion}
In this paper, we introduce an adaptation of the Rational Speech Acts (RSA) framework for Information Retrieval (IR), which we call the Rational Retrieval Acts (RRA). This framework is compatible with any sparse IR model; here, we specifically apply it to four 
 state-of-the-art sparse neural IR models and BM25. Our results demonstrate that RRA enhances models effectiveness without compromising efficiency on out-of-domain datasets. 
 A limitation is that RRA's offline RSA phase must be re-applied to the entire collection, including new documents, whenever the dataset is updated -- but approximate updates (only computing the document representations) can be used in this case.
Finally, our work also underlines that document representations, as computed by sparse neural IR models, might not be optimal at a collection level, and poses the question of computing -- especially for out-of-domain datasets -- better representations that are collection-dependent.  
\begin{acks}
The authors acknowledge Air Liquide for the financial support and ANR - FRANCE (French National Research Agency) for its financial support of the GUIDANCE project n°ANR-23-IAS1-0003. 
\end{acks}

\bibliographystyle{plainnat}
\bibliography{refs}

\appendix

\end{document}